# Exploration of Spinodal Decomposition in Multi-Principal Element Alloys (MPEAs) using CALPHAD Modeling

**Kamalnath Kadirvel*, Shalini Roy Koneru*, and Yunzhi Wang†**

Department of Materials Science and Engineering, The Ohio State University, Columbus, Ohio

## Abstract

*Researchers attributed the orderly arranged nanoscale phases observed in many multi-principal element alloys (MPEAs) to spinodal/spinodal-mediated phase transformation pathways. However, spinodal decomposition is not well understood in multicomponent systems. Although the theoretical background is available, CALPHAD databases were not used to explore the miscibility gap in MPEAs. In this work, we develop a CALPHAD framework utilizing the Hessian of free energy to study the stability of solid solutions in MPEAs. In particular, we utilize the geometry of higher dimensional Gibbs simplex in conjunction with the Hessian to calculate concentration modulations in early stages of spinodal decomposition. We apply this framework to a diverse set of multi-phase MPEAs that have been studied in the literature including TiZrNbTa (BCC/BCC), $Fe_{15}Co_{15}Ni_{20}Mn_{20}Cu_{30}$ (FCC/FCC), $Al_{0.5}NbTa_{0.8}Ti_{1.5}V_{0.2}Zr$ (BCC/B2), $AlCo_{0.4}Cr_{0.6}FeNi$ (BCC/B2) and $Al_{0.5}Cr_{0.9}FeNi_{2.5}V_{0.2}$ (FCC/L12). We show that the MPEA systems are unstable only to certain concentration modulations which could be further explored to design microstructurally engineered alloys.*

* These authors contributed equally to this work.

† Corresponding author, Email: wang.363@osu.edu

Kamalnath Kadirvel, Email: kadirvel.1@buckeyemail.osu.edu

Shalini Roy Koneru, Email: koneru.21@osu.edu

The discovery of single solid solution phases in multi-principal element alloys (MPEAs) in early 2000's has opened up a new research field commonly known as "high entropy alloys (HEAs)" [1-4]. Many MPEAs have large positive enthalpies of mixing and are likely to have a miscibility gap [3, 5-7]. The nanoscale periodic two-phase microstructures in MPEAs such as $Al_{0.5}NbTa_{0.8}Ti_{1.5}V_{0.2}Zr$ [8] and $AlCo_{0.4}Cr_{0.6}FeNi$ [9] are believed to have evolved through spinodal decomposition mediated phase transformation pathways. The presence of such uniform nanoscale two-phase microstructures in MPEAs can enhance the mechanical [10-13] and functional [9, 14] properties. Thus, a better understanding of spinodal decomposition in MPEAs can assist in tailoring the microstructure to obtain desired mechanical or functional properties. However, spinodal decomposition in MPEAs was not explored sufficiently in the literature through CALPHAD databases. Therefore, the goal of the current study is to develop a CALPHAD framework utilizing existing spinodal decomposition theories to study phase separation in MPEAs that have been explored in the literature.

Following the seminal work of Cahn [15], several researchers including Khachaturyan [16, 17], de Fontaine [18, 19] and Morral [20, 21] developed the spinodal decomposition theory in multicomponent systems. In particular, de Fontaine used a discrete free energy model [22] and developed a thermodynamic stability criteria for multicomponent systems that can be used to identify the spinodal boundary [18]. The simulated microstructures of ternary spinodal decomposition show that the final microstructure morphology is highly dependent on the alloy composition location with respect to the spinodal boundary [23-25]. The initial concentration modulations (CMs) during the early stage of spinodal decomposition can be quite different from the equilibrium tie-line [21] and could play a vital role in microstructure evolution [8]. After the discovery of MPEAs, Morral and Chen [5, 6, 26, 27] studied various aspects of spinodal decomposition in MPEAs using prototype systems, but not with CALPHAD databases. Recently, CALPHAD databases like TCHEA [28, 29] and PanHEA [30] were developed specifically for MPEAs. However, CALPHAD calculations in the literature with these databases are predominantly limited to isopleth sections and high-throughput calculation of equilibrium and metastable phases [31-41]. Thus, in this study we calculate spinodal boundaries and initial CMs from the Hessian of the free energies of the MPEAs studied in the literature.

Firstly, we review briefly de Fontaine's stability criterion in a simplified manner [18] using the continuum form of the free energy and its extension to calculate initial CMs [19, 42, 43]. A CALPHAD framework is then proposed and demonstrated on a quaternary system. Secondly, we apply the framework to a diverse set of alloys that form various two-phase microstructures including BCC/BCC (TiZrNbTa [10, 44]), FCC/FCC ($Fe_{15}Co_{15}Ni_{20}Mn_{20}Cu_{30}$ [14]), BCC/B2 ($Al_{0.5}NbTa_{0.8}Ti_{1.5}V_{0.2}Zr$ [8] and $AlCo_{0.4}Cr_{0.6}FeNi$ [9]) and FCC/L12 ($Al_{0.5}Cr_{0.9}FeNi_{2.5}V_{0.2}$ [11]). Thirdly, we compare the initial and equilibrium CMs in the alloy $Fe_{15}Co_{15}Ni_{20}Mn_{20}Cu_{30}$ and discuss its significance. Finally, we compute initial CMs for other alloys studied in this work and compare them with the experimental observations. Note that we do not consider coherent spinodal in the current work due to the lack of reliable lattice-misfit data of decomposing phases. The coherent spinodal surface will be topologically similar to the chemical spinodal with a reduced critical temperature [6, 45].

De Fontaine's analysis [18] is presented here in a simplified manner utilizing the continuum free energy formulation of Cahn [15]. Let us consider an $N$-component alloy of composition $\vec{c} = (c_1, c_2, c_3, \dots, c_{N-1})$ where the $N$-th element is chosen as the dependent element ($c_N = 1 - \sum_{i=1}^{N-1} c_i$). Assume an arbitrary fluctuation in concentration $\vec{\Delta} = (\Delta_1, \Delta_2, \dots, \Delta_{N-1})$ such that half of the system has composition $(\vec{c}+\vec{\Delta})$ and the other half has $(\vec{c}-\vec{\Delta})$, thus conserving the overall composition (Figure 1a). Note that $\Delta_N = -\sum_{i=1}^{N-1}\Delta_i$ for a physically possible fluctuation. The free energy change, $\Delta F^{sys}$, of the system per mole due to the fluctuation, $\vec{\Delta}$, is given by

$$\Delta F^{sys} = \frac{1}{2}F(\vec{c_o}+\vec{\Delta}) + \frac{1}{2}F(\vec{c_o}-\vec{\Delta}) - F(\vec{c_o}) \quad (1)$$

The Taylor expansion of the free energy, $F(\vec{c_o}+\vec{\Delta})$, up to the second order is given by

$$F(\vec{c_o}+\vec{\Delta}) = F(\vec{c_o}) + \frac{\partial F}{\partial c_i}\Delta_i + \frac{1}{2}\frac{\partial^2 F}{\partial c_i \partial c_j}\Delta_i\Delta_j + \cdots \quad (2)$$

where Einstein-convention is used for the summation. By substituting $F(\vec{c_o}+\vec{\Delta})$ and $F(\vec{c_o}-\vec{\Delta})$ in Eq. (1) using the above Taylor expansion, we get

$$\Delta F^{sys} = \Delta^T \boldsymbol{H}\,\Delta; \qquad H_{ij} = \frac{\partial^2 F}{\partial c_i \partial c_j} \quad (3)$$

Let the eigenvalues and eigenvectors of $\boldsymbol{H}$ be $\{\mu_m\}$ and $\{\overrightarrow{\Lambda^m}\}$, respectively ($m = 1,2,3,\ldots,N-1$; $\boldsymbol{H}\,\overrightarrow{\Lambda^m} = \mu_m\overrightarrow{\Lambda^m}$). The eigenvalues are labelled in ascending order ($\mu_1 \le \mu_2 \le \cdots \le \mu_{N-1}$). By resolving the fluctuation $\vec{\Delta}$ along the eigenvector directions ($\vec{\Delta} = \sum_m d^m\overrightarrow{\Lambda^m}$ where $d^m = \vec{\Delta}.\overrightarrow{\Lambda^m}$), the free energy change can be simplified as

$$\Delta F^{sys} = \sum_{m=1}^{N-1}\mu_m\,(d^m)^2 \quad (4)$$

From Eq. (4), we observe that $\Delta F^{sys}$ can be negative only if at least one of the eigenvalues $\{\mu_m\}$ is negative. In other words, the solid solution would be unstable when at least one of the eigenvalues is negative. The spinodal boundary can be identified with the condition $\mu_1 = 0$ (i.e., the smallest eigenvalue is zero). One might expect that the initial CMs will be along $\overrightarrow{\Lambda^1}$ within the spinodal region (i.e., $\mu_1 < 0$) as it maximizes the reduction in the free energy. However, as recently reported by Morral and Chen [6], the eigenvalues $\{\mu_1\}$ are not gauge invariant, i.e., the eigenvalues of the Hessian $\boldsymbol{H}$ change with the choice of the reference element (Fig. 9 of Ref. [6]). Consequently, the eigenvectors $\{\overrightarrow{\Lambda^m}\}$ also change with the choice of the reference element, and, hence, $\overrightarrow{\Lambda^1}$ cannot represent the initial CMs. Note that the spinodal boundary and $\overrightarrow{\Lambda^1}$ on the boundary are gauge invariant [6] .

The eigenvalues are not gauge invariant [6] because the basis vectors of the compositional fluctuations in Gibbs simplex are non-orthogonal (a simplex is a higher dimensional generalization of a triangle). This is schematically illustrated in Figure 1b for a ternary system, where the angle between the basis vectors $\vec{v_1}$ and $\vec{v_2}$ is $\pi/3$. The cartesian basis $X = \{\vec{x_1},\vec{x_2}\}$ can be graphically calculated in terms of the basis vectors of the Gibbs triangle (Figure 1c). Then, the orthogonalization matrix $\boldsymbol{O}$ that can be used to convert the fluctuations in the Gibbs triangle to cartesian coordinates could be computed.

$$\vec{x_1} = \vec{v_1}(-1) + \vec{v_2}(1); \;\; \vec{x_2} = \vec{v_1}\left(-\frac{1}{\sqrt{3}}\right) + \vec{v_2}\left(-\frac{1}{\sqrt{3}}\right) \quad (5)$$

$$\Delta = \boldsymbol{O}\,u; \quad \boldsymbol{O} = \begin{pmatrix} -1 & -\frac{1}{\sqrt{3}} \\ 1 & -\frac{1}{\sqrt{3}} \end{pmatrix}; \Delta = \begin{pmatrix} \Delta_1 \\ \Delta_2 \end{pmatrix}; u = \begin{pmatrix} u_1 \\ u_2 \end{pmatrix} \tag{6}$$

where $\Delta$ and $u$ are fluctuations in Gibbs and cartesian spaces respectively. We need a generalized approach to formulate the orthogonalization matrix $\boldsymbol{O}$ in higher dimensions ($N > 3$) as it is difficult to visualize the Gibbs simplex in higher dimensions. Singh et.al [42] showed that this orthogonalization matrix can be written in terms of the $(N-1)$-dimensional simplex vertices and presented an iterative algorithm to calculate the simplex vertices. Using this algorithm, they obtained the orthogonalization matrix $\boldsymbol{O}$ for a 4-dimensional simplex. In our work, we used this approach to compute $\boldsymbol{O}$ for a 9-dimensional simplex, which is shown as $\boldsymbol{T}$ in Table 1. The orthogonalization matrix $\boldsymbol{O}$ for any *N*-element system ($N \leq 10$) can be calculated from $\boldsymbol{T}$ by considering the subset square matrix of order *N-1* formed by the elements in the first *N-1* rows and *N-1* columns. In cartesian coordinates, the free energy change $\Delta F^{sys}$ can be written as

$$\Delta F^{sys} = u^T\,\widehat{\boldsymbol{H}}\,u; \quad \widehat{\boldsymbol{H}} = \boldsymbol{O}^{\boldsymbol{T}}\boldsymbol{H}\,\boldsymbol{O} \tag{7}$$

where $\widehat{\boldsymbol{H}}$ is the Hessian in cartesian basis. The eigenvectors $\{\overrightarrow{U^m}\}$ and eigenvalues $\{\lambda_m\}$ of $\widehat{\boldsymbol{H}}$ are gauge-invariant and are used in our analysis $(m = 1,2,3,\ldots,N-1)$. The eigenvectors $\{\overrightarrow{U^m}\}$ from the cartesian space are transformed back into the Gibbs space using $\overrightarrow{V^m} = \boldsymbol{O}\overrightarrow{U^m}$. Based on the eigenvalues $\{\lambda_m\}$, a multicomponent solid solution would be

a) unstable to all CMs if all the eigenvalues are negative,

b) unstable to only certain CMs if some of the eigenvalues are negative, or

c) absolutely stable if all the eigenvalues are positive

The detailed steps of our CALPHAD framework and the illustrative result of our stability analysis are shown in Figure 1. We extract the Hessian matrix of the parent solid solution phase from Thermo-Calc [46] (using TCHEA4 database) and perform eigenvalue/eigenvector analysis in MATLAB [47]. The eigenvalues and eigenvectors are evaluated in 400–1400 K temperature interval in steps of 40 K. As there is no inherent ordering in the eigenvalues, we index them in ascending order to maintain consistency (i.e., $\lambda_1 \leq \lambda_2 \leq \cdots \leq \lambda_{N-1}$). The eigenvector component of the reference element is calculated as $V_N^m = -\sum_i^{N-1} V_i^m$. To demonstrate the method, the calculated eigenvalues $(\lambda_1, \lambda_2, \lambda_3)$ and the eigenvector $\overrightarrow{V^1}$ of a quaternary alloy TiZrNbTa (BCC) are shown in Figure 2b and Figure 2c, respectively. Only one of the eigenvalues is negative at lower temperatures, which implies that TiZrNbTa would be *unstable* to only certain initial CMs that can grow at lower temperatures. The direction in the compositional space (initial CMs) along which the system has maximum instability (i.e., maximum reduction in free energy $\Delta F^{sys}$) is given by the eigenvector $\vec{V}^1$ corresponding to the negative eigenvalue (i.e., $\lambda_1$). Elements with the same sign eigenvector components (i.e., positive, or negative) will partition together in the early stage of the spinodal decomposition. If the magnitude of certain elemental component is less than 5% of the largest eigenvector component, then the elemental partitioning of those elements is considered negligible in our work. For instance, from the eigenvector components (Figure 2c), we expect the formation of Nb–Ta-rich and Ti–Zr-rich phases. The current predictions agree well with the recent experimental observations of nanoscale interconnected Ti–Zr-rich and Nb–Ta-rich phases that occur during continuous cooling of a homogenized TiZrNbTa alloy [10, 44].

Such stability analysis is performed for the MPEAs listed in Table 2 to verify and explore the experimentally proclaimed hypothesis of spinodal and spinodal-mediated phase transformations [8-11, 14]. The experimentally observed phases, the parent phase chosen for the stability analysis, the initial elemental partitioning and the calculated spinodal temperature are also listed in the table. In the temperature range of our analysis, at most one eigenvalue is found to be negative for all the MPEAs. Hence, we defined a stability factor (SF) for each alloy based on its minimum eigenvalue (i.e., $\lambda_1$).

$$SF = \frac{\lambda_1}{\lambda_1^{ideal}} \tag{8}$$

where $\lambda_1$ is calculated from the database (Figure 2a) and $\lambda_1^{ideal}$ is calculated by assuming that the alloy is an ideal solid solution. For a near ideal solution, $SF \approx 1$. In the spinodal region, $SF < 0$ and at the spinodal boundary $SF = 0$. SF as a function of temperature for all the alloys in Table 2 is shown in Figure 3. The Cantor alloy [3] is one of the well-studied FCC solid solution with no miscibility gap. Hence, we used this alloy as a negative-test case, and, as expected, the SF is positive for all temperatures implying that the solid solution is *absolutely stable* with respect to spinodal decomposition. In agreement with the experimental observations, the alloys $Fe_{15}Co_{15}Ni_{20}Mn_{20}Cu_{30}$ [14], $Al_{0.5}NbTa_{0.8}Ti_{1.5}V_{0.2}Zr$ [8], and $AlCo_{0.4}Cr_{0.6}FeNi$ [9] exhibit negative SF at lower temperatures in the explored temperature range. Note that the cusp in the $AlCo_{0.4}Cr_{0.6}FeNi$ curve around 650 K is likely due to the magnetic transition of the alloy (the calculated Curie temperature for this alloy is 652 K). For the alloy $Al_{0.5}Cr_{0.9}FeNi_{2.5}V_{0.2}$, the calculated spinodal temperature is 474 K. Experimentally, spinodal decomposition was indeed observed in this alloy but at temperatures as high as 973 K [11]. It seems that further optimization of the CALPHAD thermodynamic databases is necessary to make more accurate predictions.

The calculated element partitioning from the eigenvectors along with the experimentally observed partitioning is described in Table 2 for all the alloys examined in this study. Note that the experimentally measured elemental partitioning would be closer to the initial CMs rather equilibrium compositions as the experimental microstructures were either formed during continuous cooling or short-time isothermal annealing. The processing conditions of the alloys are listed in Table S1 (Supplementary material).

In the alloy $Fe_{15}Co_{15}Ni_{20}Mn_{20}Cu_{30}$ undergoing spinodal decomposition, the equilibrium CMs ($\vec{E}$) could be calculated as

$$\vec{E} = \frac{\vec{c}(FCC\#1) - \vec{c}(FCC\#2)}{d} \tag{9}$$

where the equilibrium compositions $\vec{c}(FCC\#1)$ and $\vec{c}(FCC\#2)$ are calculated from the database and the normalization factor $d$ is calculated such that $E_{Cu} = V_{Cu}^1$. For $Fe_{15}Co_{15}Ni_{20}Mn_{20}Cu_{30}$, the eigenvector components in Figure 4a suggests that (i) Cu will partition out from Fe and Co and (ii) Ni and Mn have negligible partitioning in the early stage of spinodal decomposition. The comparison between the initial and equilibrium CMs at 873 K is shown in Figure 4b. Interestingly, even though the initial CMs in Ni and Mn are negligible, the computed equilibrium CMs show a clear partitioning behaviour in Ni and Mn comparable to that of Fe and Co. A recent experimental observation of spinodal decomposition in this alloy reveals Cu-rich and Fe–Co-rich regions after annealing for 6 h at T = 873 K, as shown in Figure 4c–f [14]. Specifically, the EDS maps in Figure 4f show negligible partitioning of Ni and Mn in comparison to other elements like Fe, Co, and Cu. This highlights the need to consider initial CM in addition to equilibrium compositions in microstructural design of MPEAs. Note that in the experimental study [14], negligible partitioning in Ni and Mn could not be explained as only the spinodal temperature and the equilibrium compositions were calculated while the initial CMs were not analysed.

Finally, we study the initial CMs in BCC/B2 and FCC/L12 alloys listed in Table 2. The spinodal mediated transformation pathways through which the BCC/B2 or FCC/L12 microstructures could have evolved were already explored in the literature [48, 49]. For the case of $Al_{0.5}NbTa_{0.8}Ti_{1.5}V_{0.2}Zr$ (Figure 5a), the eigenvector predicts the formation of Nb–Ta–V-rich and Al–Zr–Ti-rich phases, in agreement with the short-time ageing experiments (0.5 h at 873 K) [8]. Note that the partitioning behaviour of Ti changes during long-time ageing (120 h at 873 K) and the microstructure exhibits Nb–Ta–V–Ti-rich BCC and Al–Zr-rich B2 phases. However, this change in partitioning behaviour was not predicted by the database (Table S2 of Supplementary material). For $AlCo_{0.4}Cr_{0.6}FeNi$, the eigenvectors (Figure 5b) predict the formation of Al–Ni-rich and Cr–Fe–Co-rich phases. However, the recent experiment [9] exhibits Al–Ni–Co-rich and Cr–Fe-rich phases, contradicting the predicted co-partitioning. The observed discrepancies in element partitioning of BCC/B2 alloys highlight the well-known need to improve the B2 phase description in the CALPHAD databases [7, 13]. For the case of $Al_{0.5}Cr_{0.9}FeNi_{2.5}V_{0.2}$ (FCC/L12 alloy), the eigenvector (Figure 5c) suggests (i) the formation of Al–Ni-rich and Fe–Cr-rich phases and (ii) negligible partitioning in V. Both these predictions agree well with the experiments [11], although our stability analysis predicts lower spinodal temperatures as discussed earlier.

In summary, we established a CALPHAD framework to predict the stability of solid solutions in various MPEA systems against spinodal decomposition. Our calculations indicate that most MPEAs reported in the literature undergoing spinodal decomposition are only unstable with respect to certain CMs, which are preferred in early stages of decomposition. Obviously, the predictions based on our stability analysis are only as good as the available CALPHAD thermodynamic databases that were usually optimized using experimental data of equilibrium compositions. Our work suggests that the stability analysis along with experimental measurements of concentrations modulations at early stages of decomposition could serve as an efficient tool to refine and improve the CALPHAD databases for MPEAs. The framework established will aid studies on solid solution stability in multicomponent compositional and temperature space for MPEAs and could potentially accelerate the design of either single-phase MPEAs or multi-phase MPEAs with desired microstructures.

The authors gratefully acknowledge many helpful discussions with Dr. Shuanglin Chen from CompuTherm LLC and Professor T.A. Abinandanan at Indian Institute of Science. The work was supported by Air Force Office of Scientific Research (AFOSR) under grant FA9550-20-1-0015.

## Data Availability

The data that supports the findings of this study are available from the corresponding author upon request.

# Figures

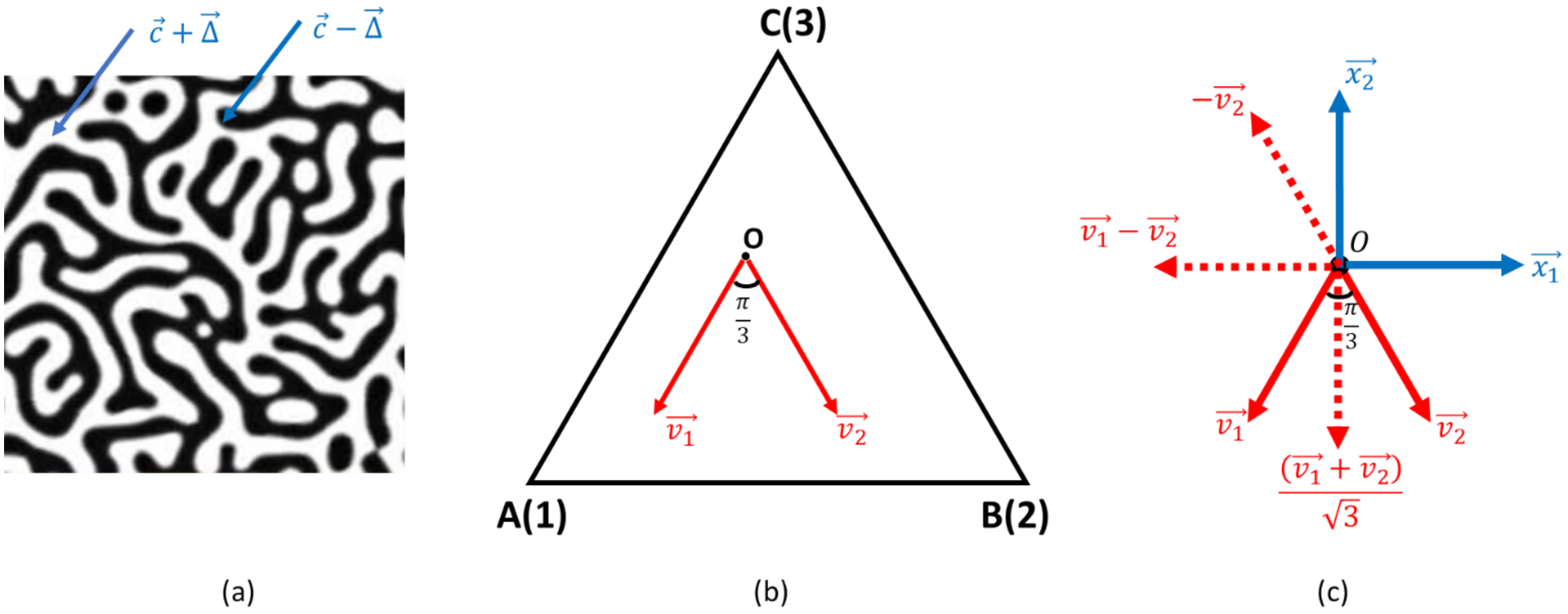

Figure 1. (a) Schematic microstructure of alloy $\vec{c}$ with fluctuation $\vec{\Delta}$ showing the regions with the concentration $(\vec{c}+\vec{\Delta})$ and $(\vec{c}-\vec{\Delta})$, respectively. (b) Gibbs triangle (A-B-C): the alloy composition $\vec{c}=(c_1,c_2)$ is marked as point **O**. Assuming "C" to be dependent element, the fluctuation $\vec{\Delta}$ can be written in terms of unit vectors $\vec{v_1}$ and $\vec{v_2}$ as $\vec{\Delta}=(\Delta_1,\Delta_2)=\Delta_1\vec{v_1}+\Delta_2\vec{v_2}$ . (c) Orthogonalization: $X=\{\vec{x_1},\vec{x_2}\}$ is the orthonormal basis set for the fluctuations that can be computed graphically $(\vec{x_1}=-(\vec{v_1}-\vec{v_2});\ \vec{x_2}=-(\vec{v_1}+\vec{v_2})/\sqrt{3}\ )$

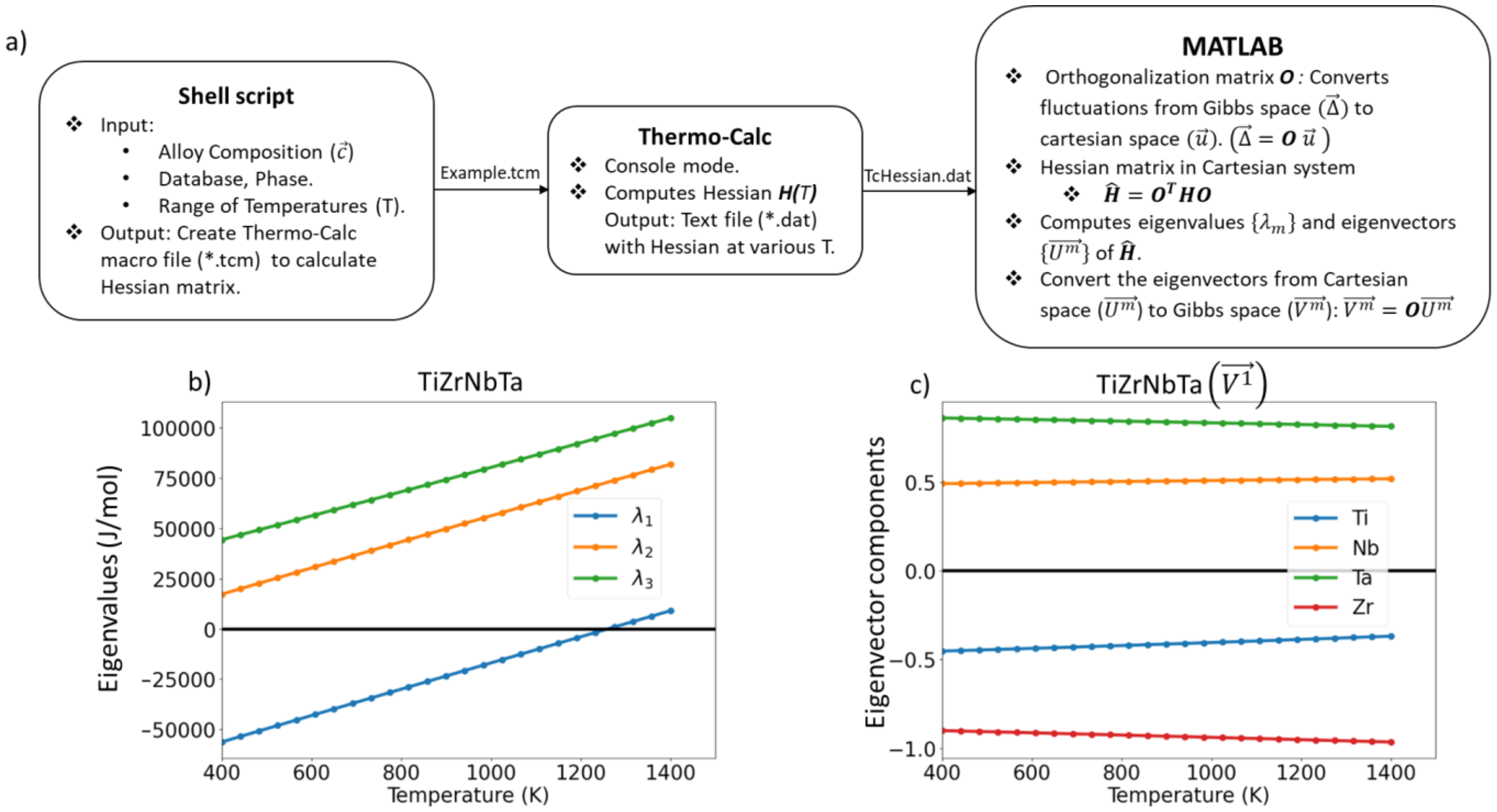

Figure 2. (a) Steps involved in the stability analysis of multicomponent solid solution using CALPHAD database. (b) Eigenvalues and (c) eigenvector components corresponding to the eigenvalue ($\lambda_1$) as a function of temperature for TiZrNbTa.

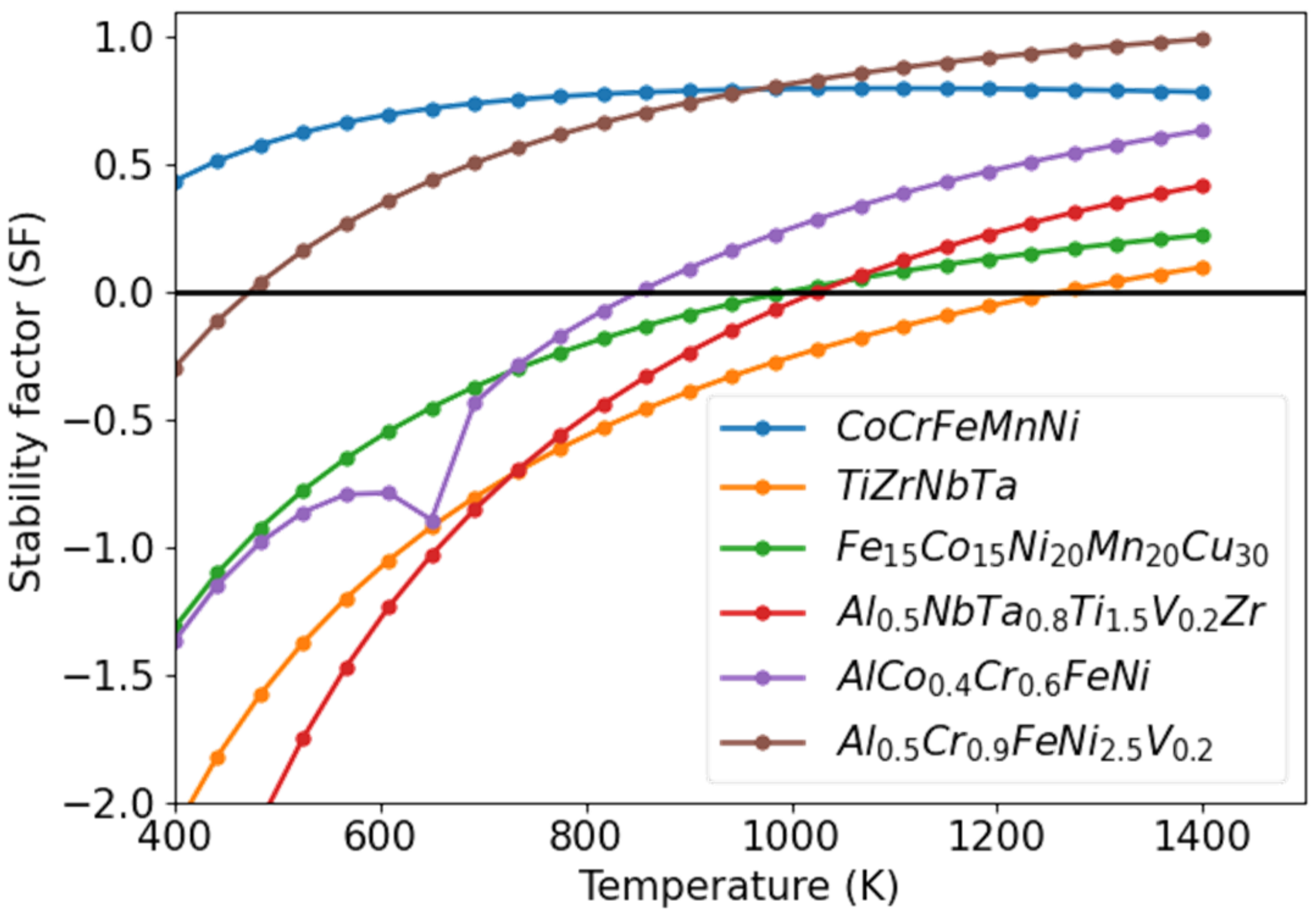

Figure 3. Stability factor (SF) as a function of temperature for the MPEAs considered in this study.

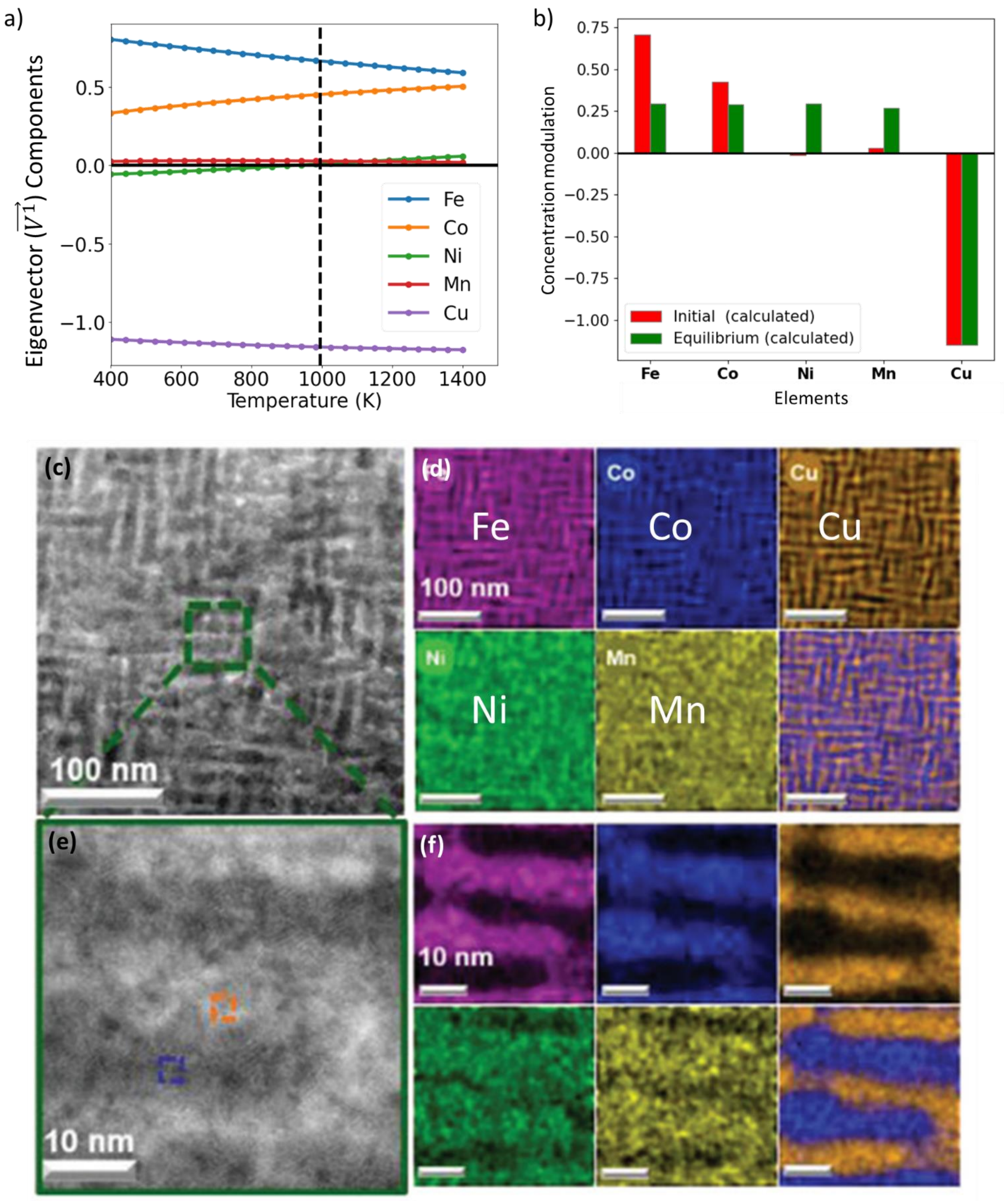

Figure 4. Analysis of the $Fe_{15}Co_{15}Ni_{20}Mn_{20}Cu_{30}$ alloy: (a) Eigenvector components $\overrightarrow{V^1}$ as a function of temperature (b) Calculated initial and equilibrium concentration modulation from our framework at T=873 K, which are quite different from each other for most of the elements. (c-f) Experimental observations of spinodal

decomposition in $Fe_{15}Co_{15}Ni_{20}Mn_{20}Cu_{30}$ at 873 K for 6 h (reproduced from [14] with permission). (c) Low magnification HAADF-STEM image. (d) EDS map of local composition of Fe, Co, Cu, Ni, Mn and the combined concentration shown in (c). (e) High magnification image of the green box in (c). (f) High magnification image of the EDS composition map (of the area marked in green box).

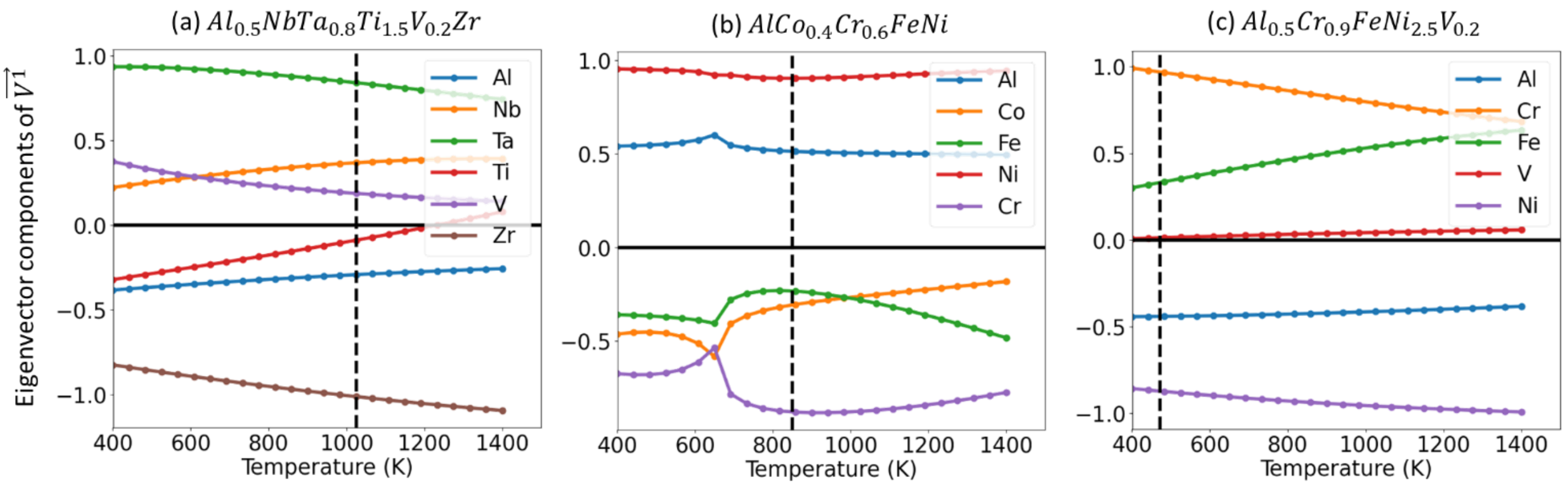

Figure 5. Eigenvector components of $\overrightarrow{V^1}$ as a function of temperature for various MPEAs. (a) $Al_{0.5}NbTa_{0.8}Ti_{1.5}V_{0.2}Zr$, (b) $AlCo_{0.4}Cr_{0.6}FeNi$, (c) $Al_{0.5}Cr_{0.9}FeNi_{2.5}V_{0.2}$. The vertical black dotted line indicates the spinodal temperature.

## Tables

Table 1. Orthogonalization matrix $\boldsymbol{T}$ $(9 \times 9)$ for 10-component system. The orthogonalization matrix $\boldsymbol{O}$ for $N$-component system $(N \leq 10)$ can be calculated from $\boldsymbol{T}$ : $O_{ij} = T_{ij}$ $(i, j = 1,2, \dots, N-1)$.

$$\boldsymbol{T} = \begin{vmatrix} -1 & -\frac{1}{\sqrt{3}} & -\frac{1}{\sqrt{6}} & -\frac{1}{\sqrt{10}} & -\frac{1}{\sqrt{15}} & -\frac{1}{\sqrt{21}} & -\frac{1}{2\sqrt{7}} & -\frac{1}{6} & -\frac{1}{3\sqrt{5}} \\ 1 & -\frac{1}{\sqrt{3}} & -\frac{1}{\sqrt{6}} & -\frac{1}{\sqrt{10}} & -\frac{1}{\sqrt{15}} & -\frac{1}{\sqrt{21}} & -\frac{1}{2\sqrt{7}} & -\frac{1}{6} & -\frac{1}{3\sqrt{5}} \\ 0 & \frac{2}{\sqrt{3}} & -\frac{1}{\sqrt{6}} & -\frac{1}{\sqrt{10}} & -\frac{1}{\sqrt{15}} & -\frac{1}{\sqrt{21}} & -\frac{1}{2\sqrt{7}} & -\frac{1}{6} & -\frac{1}{3\sqrt{5}} \\ 0 & 0 & \frac{\sqrt{3}}{\sqrt{2}} & -\frac{1}{\sqrt{10}} & -\frac{1}{\sqrt{15}} & -\frac{1}{\sqrt{21}} & -\frac{1}{2\sqrt{7}} & -\frac{1}{6} & -\frac{1}{3\sqrt{5}} \\ 0 & 0 & 0 & \frac{2\sqrt{2}}{\sqrt{5}} & -\frac{1}{\sqrt{15}} & -\frac{1}{\sqrt{21}} & -\frac{1}{2\sqrt{7}} & -\frac{1}{6} & -\frac{1}{3\sqrt{5}} \\ 0 & 0 & 0 & 0 & \frac{\sqrt{5}}{\sqrt{3}} & -\frac{1}{\sqrt{21}} & -\frac{1}{2\sqrt{7}} & -\frac{1}{6} & -\frac{1}{3\sqrt{5}} \\ 0 & 0 & 0 & 0 & 0 & \frac{2\sqrt{3}}{\sqrt{7}} & -\frac{1}{2\sqrt{7}} & -\frac{1}{6} & -\frac{1}{3\sqrt{5}} \\ 0 & 0 & 0 & 0 & 0 & 0 & \frac{\sqrt{7}}{2} & -\frac{1}{6} & -\frac{1}{3\sqrt{5}} \\ 0 & 0 & 0 & 0 & 0 & 0 & 0 & \frac{4}{3} & -\frac{1}{3\sqrt{5}} \end{vmatrix}$$

Table 2: Alloys analysed in our work. The element partition columns (both experimental and calculated) correspond to the initial concentration modulations (CMs). *Calculated elemental partitioning is considered negligible if the corresponding eigenvector component is less than 5% of the largest eigenvector component.

| **No.** | **Alloy** | **Phases present in the bulk** | **Element partition (Experimental)** | | **Parent phase (in CALPHAD calculation)** | **Element partition (calculated)** | | $T_{spi}$ **(K)** |
|---|---|---|---|---|---|---|---|---|
| | | | **Phase1** | **Phase2** | | **Phase1** | **Phase2** | |
| 1 | CoCrFeMnNi (Cantor alloy) [3] | FCC | - | - | FCC | - | - | - |
| 2 | TiZrNbTa [10] | BCC+BCC | Ti, Zr | Nb, Ta | BCC | Ti, Zr | Nb, Ta | 1260 |
| 3 | $Fe_{15}Co_{15}Ni_{20}Mn_{20}Cu_{30}$ [14] | FCC+FCC | Cu | Fe, Co (Ni and Mn negligible) | FCC | Cu | Fe, Co (Ni and Mn negligible *) | 996 |
| 4 | $Al_{0.5}NbTa_{0.8}Ti_{1.5}V_{0.2}Zr$ [8] | BCC+B2 | Al, Zr, Ti | V, Nb, Ta | BCC | Al, Zr, Ti | V, Nb, Ta | 1025 |
| 5 | $AlCo_{0.4}Cr_{0.6}FeNi$ [9] | BCC+B2 | Al, Ni, Co | Fe, Cr | BCC | Ni, Al | Fe, Co, Cr | 851 |
| 6 | $Al_{0.5}Cr_{0.9}FeNi_{2.5}V_{0.2}$ [11] | FCC+L12 | Al, Ni | Fe, Cr (V negligible) | FCC | Ni, Al | Fe, Cr (V negligible *) | 474 |